\documentclass[letterpaper]{article} 
\usepackage{aaai2027}
\usepackage[hyphens]{url}  
\usepackage{graphicx} 
\usepackage{natbib}  
\usepackage{caption} 
\usepackage{amsmath}
\usepackage{amssymb}
\usepackage{amsthm}
\usepackage{algorithm}
\usepackage{algorithmic}
\usepackage{booktabs}
\usepackage{fancyvrb}
\usepackage{tabularx}
\usepackage{fvextra}
\usepackage{xcolor}

\title{REFINE: A Resilient Evolution Framework for Intelligent Enterprise Alert Triage in Security Operations Centers}
\author {
    Huimin Chen\textsuperscript{\rm 1},
    Quan Long\textsuperscript{\rm 1}\corresponding,
    Yanhao Wang\textsuperscript{\rm 2}
}
\affiliations {
    \textsuperscript{\rm 1}Baidu Inc.\\
    \textsuperscript{\rm 2}East China Normal University\\
}

\begin{document}

\nocopyright
\maketitle

\begin{abstract}
Security Operations Centers (SOCs) process large volumes of alerts daily.
Alert triage is a core operation in SOCs, used to classify alerts so that analysts can focus less on low-risk alerts while never missing high-risk alerts requiring human intervention.
Recently, LLM agents show promise in reasoning jointly over logs, assets, threat intelligence, and historical dispositions, but applying them directly to alert triage still fails to meet operational requirements.
The root cause is that alert triage depends heavily on organization-specific, fast-evolving asset priorities, business context, historical disposition norms, and analyst risk preferences, making static prompts, rules, or even expert experience unable to remain aligned with operational standards over time.

In this paper, we present \textbf{REFINE}, a new \textbf{R}esilient \textbf{E}volution \textbf{F}ramework for \textbf{IN}telligent \textbf{E}nterprise alert triage using LLM agents.
REFINE encodes analyst triage expertise as a \emph{skill} that jointly specifies judgment rules, evidence-chain reasoning, and tool invocation protocols, and continually discovers divergences between the skill and current operational standards through real disposition feedback.
Furthermore, it incorporates high-risk alert recall as a hard constraint in the optimization by enforcing $\text{recall}=1.0$ on the evolution set to expand the range of false positives the agent can auto-close; it also locates judgment blind spots by combining the global alert distribution with the current skill's error boundary, focusing feedback utilization on misjudgment patterns that affect the safety baseline in real deployment.
We evaluate REFINE on four alert scenarios spanning four MITRE ATT\&CK phases in a real industrial SOC.
On the evolution sets, REFINE is the only method reaching $\text{recall}=1.0$ in all four scenarios, and it uses this feasible region to expand benign false-positive auto-closure.
On the later test windows, REFINE still maintains $\text{recall}=1.0$ in three of the four scenarios; in the one scenario where its recall does drop to $0.807$, it is still higher than those of the evaluated self-evolution baselines ($0.49$--$0.58$).
\end{abstract}


\section{Introduction}

Enterprise Security Operations Centers (SOCs) receive alert volumes far beyond human triage capacity, and alert fatigue has severely weakened threat detection~\cite{tariq2025alert,veeramachaneni2016ai}.
Consequently, automatic alert triage that closes large volumes of benign false positives to reduce analyst workload while avoiding missing threats that require human intervention is urgently needed.
In addition, as the costs of false positives and negatives are highly asymmetric, i.e., a false positive that is not closed merely remains in the human queue, whereas a missed recall may leave an ongoing intrusion unhandled, expanding false-positive auto-closure under a no-missed-recall safety baseline is, therefore, a central operational problem for SOCs.

Recently, Large Language Model (LLM) agents show immense potential in jointly reasoning over logs, assets, threat intelligence, and historical dispositions, paving a new way for alert triage in SOCs~\cite{ferrag2024revolutionizing,hasanov2024application}.
However, general LLM reasoning cannot fully reproduce an enterprise's actual triage behavior because SOC triage depends on an organization-specific \emph{operational scale}: asset priorities, business context, historical disposition norms, and risk tolerance~\cite{vielberth2020security,moosmann2026can}.
This scale includes analyst expertise that is difficult to articulate completely~\cite{polanyi2009tacit}, may be fragmented by differences among analysts, and evolves with business context and the threat landscape; even when encoded as rules, a skill's design intent may diverge from the agent's actual execution boundary.
Consequently, pretrained models, manual rule sets, or static skills cannot remain aligned with operational standards over time.

Existing studies, however, do not fully address such cross-domain problems.
Conventional SOC methods mainly train static classifiers~\cite{ban2021combat} or adjust alert rankings~\cite{jalalvand2024alert}, making it difficult to update organization-specific triage specifications based on newly arriving disposition feedback.
General self-evolution methods such as GEPA~\cite{agrawal2026gepa} and ACE~\cite{zhang2025agentic} can optimize policies through reflection, but they target general task objectives rather than two key SOC requirements: continuously aligning organizational scale from real alerts and disposition feedback, and preserving the no-missed-recall safety baseline while improving auto-closure.
What SOCs require is thus not a general-purpose skill evolution scheme but a continuous alignment mechanism that integrates security operational knowledge, agent feedback learning, and asymmetric safety constraints.

In this paper, we present a \underline{R}esilient \underline{E}volution \underline{F}ramework for \underline{IN}telligent \underline{E}nterprise alert triage (\textbf{REFINE}) using LLM agents.
REFINE encodes analyst triage expertise as a \emph{skill}---a structured agent behavior specification that jointly defines judgment rules, evidence-chain reasoning, and tool invocation protocols---and uses real alert dispositions to continuously identify divergences between the skill and the current operational scale.
It combines the global alert distribution with the current skill's error boundary to localize blind spots, prioritizing limited feedback toward error patterns that matter in real operations.
It further incorporates high-risk alert recall as a hard constraint and uses multi-round reflection with a recall gate to expand benign false-positive auto-closure without missing recalls.
By distilling analysts' tacit, fragmented triage expertise into a single, maintainable, and continuously evolvable explicit skill, REFINE converges fragmented individual scales into a unified organization-level triage standard.

Experiments on four alert scenarios spanning four MITRE ATT\&CK phases in a real industrial SOC separate evolution-set and test-window behavior.
On the evolution sets, REFINE is the only method reaching $\text{recall}=1.0$ in all four scenarios while substantially expanding benign-alert closure (e.g., NTLM-Hash evolution set accuracy rising from $0.217$ to $0.907$ at $\text{recall}=1.0$).
On the strictly later test windows, it maintains $\text{recall}=1.0$ in three of the four scenarios; in the one scenario where recall drops, its $0.807$ recall is still the highest among all evaluated methods (versus $0.49$--$0.58$ for other self-evolution baselines).

\section{Related Work}

\subsection{SOC Alert Triage}

Alert fatigue has been a core challenge in SOC operations, and cognitive overload directly degrades triage quality~\cite{tariq2025alert,dykstra2018cyber}.
LLMs have been widely introduced into triage workflows to alleviate staffing bottlenecks ~\cite{ferrag2024revolutionizing,hasanov2024application,ndichu2026ai,habibzadeh2025large}, and \citet{xi2025rise} have surveyed AI-driven screening methods and LLM integration across the full SOC pipeline.
However, the non-static nature of triage criteria represents a more fundamental challenge ~\cite{moosmann2026can,ban2021combat,jalalvand2024alert}.
\citet{singh2025llms} found that analysts primarily use LLMs for comprehension assistance rather than direct triage decisions;
\citet{kokulu2019matched} conducted interview studies revealing inherent disagreements between SOC managers and analysts that, if left unaddressed, undermine SOC effectiveness.
\citet{wei2025cortex} proposed CORTEX, a multi-agent collaborative architecture that improves coordination efficiency through specialized agent division of labor, but its triage strategy is static and cannot automatically update as organizational judgment criteria evolve.
The above studies reveal the pervasiveness of operational judgment discrepancies in practice; yet, all address this inherently dynamic problem with static policies.

\subsection{Self-Evolution Methods}

The self-evolution direction aims to automatically optimize natural-language specifications that guide LLM/Agent behavior, replacing manual iteration: early works ~\cite{zhou2022large,yang2024large} established foundations via candidate selection and natural-language iteration, after which evolutionary algorithms were introduced~\cite{guo2024connecting,fernando2024promptbreeder};
GEPA~\cite{agrawal2026gepa} surpasses Reinforcement Learning (RL) baselines by combining a genetic Pareto mechanism with reflective iteration;
\citet{pryzant2023automatic} and TextGrad~\cite{yuksekgonul2024textgrad} achieve gradient-style optimization through textual gradients combined with beam search and by extending automatic differentiation to the text domain, respectively;
DSPy~\cite{khattab2024dspy} compiles LLM calls into automatically tunable pipelines;
Self-Refine~\cite{madaan2023self} and Reflexion~\cite{shinn2023reflexion} achieve parameter-free improvement via self-feedback and linguistic reinforcement signals;
ACE~\cite{zhang2025agentic} treats in-context policies as evolvable rule sets;
AutoSkill~\cite{yang2026autoskill} and \citet{zhao2025agentic} study skill self-evolution and autonomous discovery, respectively.
To the best of our knowledge, none of the above studies can provide a mechanism for enforcing hard recall constraints across the entire alert corpus, making it difficult to systematically control the risk of missed detections in real-world SOC deployments, where a single missed alert can allow adversaries to persist undetected, move laterally, and cause irreversible damage.

\section{Method}

\subsection{Problem Formulation}

In security operations, all alerts are, by default, triaged by human analysts; we aim to learn a triage \emph{skill} $S$ such that a security operations agent guided by $S$ can automatically dismiss safely closable false positives.
A skill is a structured natural-language document encoding the decision procedure, judgment conditions, reasoning logic, and tool invocation conditions; its typical Markdown structure contains four sections: core principles, triage strategy, empirical lessons, and edge cases, with tool invocation conditions distributed across them (predominantly in the triage strategy section).
Under skill $S$, the agent's triage is $\hat{y} = \operatorname{agent}(x \mid S) \in \{0,1\}$, where $\hat{y}=0$ means ``suggest dismiss'' and $\hat{y}=1$ means ``suggest escalate.''
The business objective is to maximize accuracy subject to a hard recall constraint:
\begin{equation}
    \max_{S} \; \mathrm{Accuracy}(S) \quad \text{s.t.} \quad \mathrm{Recall}(S) = 1.0
\end{equation}
This asymmetry is fundamental: a missed threat (FN) means an ongoing intrusion goes undetected; an unclosed false positive (FP) merely stays in the human queue.
Key notation and threat model are given in the supplementary material.

\begin{figure*}[!t]
\centering
\includegraphics[width=0.7\textwidth]{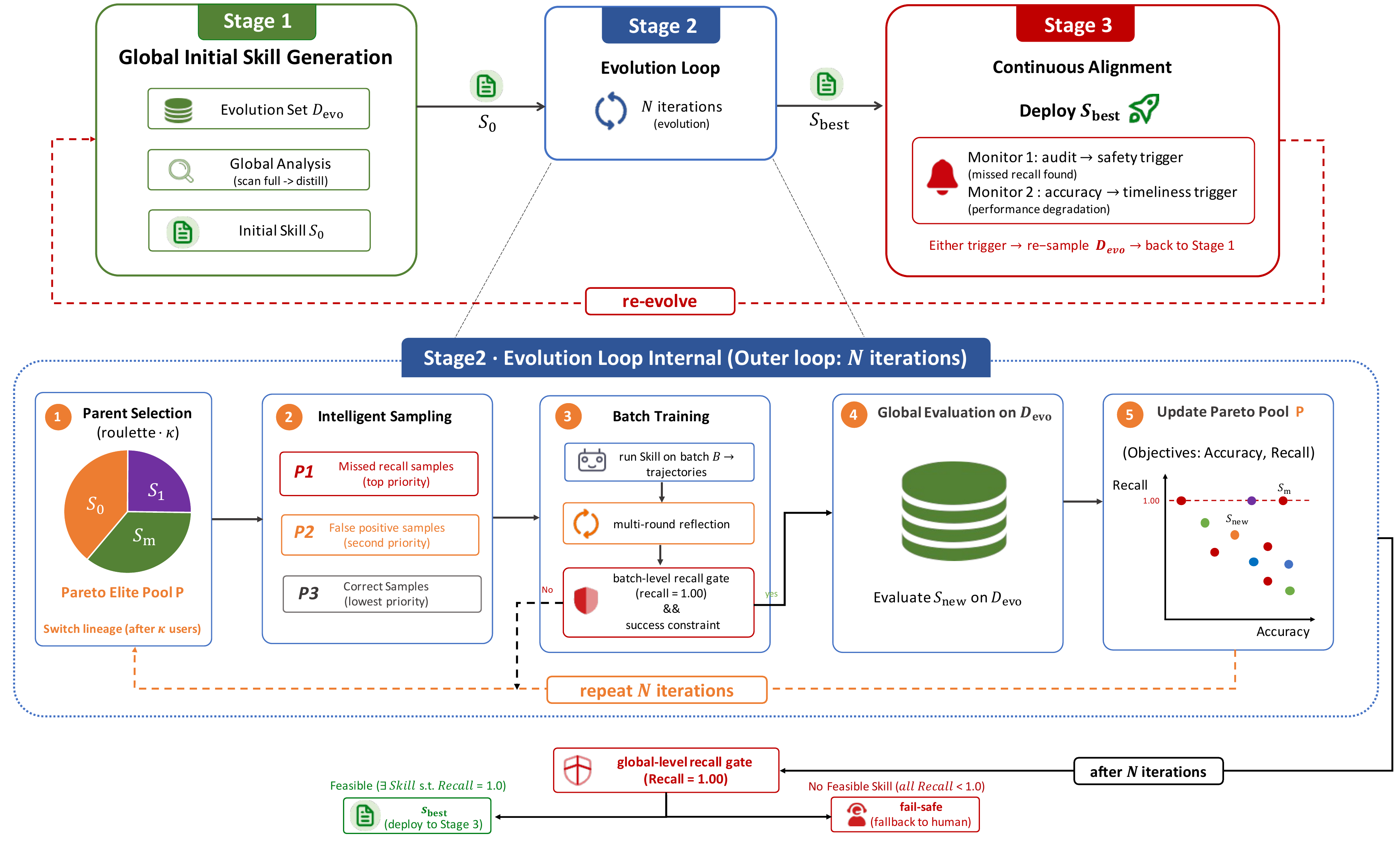}
\caption{Overall REFINE architecture. Stage 1 performs a global analysis of the evolution set $\mathcal{D}_{\text{evo}}$ to generate the initial skill $S_0$. Stage 2 improves the skill through intelligent sampling, multi-round reflection, and batch/global recall gates, then outputs the highest-accuracy feasible skill $S_{\text{best}}$ with $\mathrm{Recall}(\mathcal{D}_{\text{evo}})=1.0$ or falls back to human triage if none exists. Stage 3 deploys $S_{\text{best}}$ and triggers re-evolution from the latest production window when sampled safety checks or online accuracy indicate drift. Red components denote the recall-safety path.}
\label{fig:framework}
\end{figure*}

\subsection{REFINE Framework}

As illustrated in Figure~\ref{fig:framework}, REFINE is a skill self-evolution framework for SOC triage. Before the evolution loop begins, the agent performs a single global analysis over the full evolution set $\mathcal{D}_{\text{evo}}$ (its construction is detailed in the experimental setup) to produce an initial skill $S_0$ with full-distribution coverage; REFINE then iteratively evolves $S_0$ under the recall hard constraint through intelligent sampling and the evolution loop.
The evolution trajectory spans the entire process; post-deployment, sampled checks over auto-closed alerts and online-accuracy monitoring drive rolling self-evolution whenever the skill drifts from the current operational scale.

\paragraph{Global Initial Skill Generation}
\label{sec:global}

Typical self-evolution frameworks, e.g., GEPA \cite{agrawal2026gepa} and ACE \cite{zhang2025agentic}, evolve from a seed policy---either hand-authored or empty---and provide no automated way to construct the initial policy.
Both cases have structural limitations: an empty seed leaves the start under-informed, forcing early iterations to spend budget rediscovering the data distribution; a hand-authored seed faces a quality-cost dilemma, i.e., a casually written seed is detached from the deployment's data distribution, while a carefully written seed contradicts the automation goal of self-evolution.
Any manual bias in the seed propagates down the evolution lineage.

REFINE instead performs, before the evolution loop, a single global analysis over the complete evolution set $\mathcal{D}_{\text{evo}}$, automatically distilling the operational scale and generating an initial skill $S_0$ with full-distribution coverage, without a manual seed.
This is a one-shot summarizing pass over the full set (statistics over distribution and boundary samples, not per-sample triage) rather than an iterative procedure.
Starting from global coverage, subsequent evolution focuses on repairing known defects rather than compensating for information gaps.

\paragraph{Intelligent Sampling}
\label{sec:sampling}

Given parent skill $S$, samples are classified by security impact into three priorities:
\textbf{P1} (missed recalls, $f_S(x){=}0 \land y{=}1$),
\textbf{P2} (false positives, $f_S(x){=}1 \land y{=}0$),
and \textbf{P3} (correct).
This mirrors uncertainty sampling from active learning: prioritize the most informative samples under a limited reflection budget.
Note that batching is not a context-capacity limitation (the initial-generation step above shows the full evolution set can be ingested in a single analysis); the batched reflection input follows GEPA's evolutionary search paradigm, focusing each reflection round on a set of related errors.
REFINE departs from prior local-view methods in that both \emph{generation and evaluation are globally anchored}, with initial generation over the full set and candidate evaluation on the complete $\mathcal{D}_{\text{evo}}$, rather than in how the reflection input is batched.
P1 priority, multi-round reflection (below), and the recall gate together continuously confine the search to the recall-feasible region.

\paragraph{Evolution Loop}
\label{sec:batch}

The evolution loop is nested: an outer loop of $N$ iterations, each drawing a parent skill $S$ from the elite pool and a batch $B$ by intelligent sampling, wraps an inner loop of up to $R$ reflection rounds over $B$. In a round, the agent executes $S$ on $B$ to produce reasoning trajectories, and a reflection LLM revises $S$ into $S'$ from them plus ``what was wrong / why'' feedback (prompt template in the supplementary material).

Multiple rounds are needed because reducing FPs and avoiding new FNs compete: a single revision often improves one metric while degrading the other.
Each round therefore takes the previous round's $S'$ as its parent, extending one lineage in depth, and its output is accepted only when the \textbf{success constraint} holds against the best values reached so far in this batch, i.e., batch recall does not drop and at least one metric strictly improves, preventing within-batch regression.

When the inner loop ends, the \textbf{batch-level recall gate} returns, among the skills it produced, the one with batch $\text{recall} = 1.0$ and the highest batch accuracy; if none exists, the batch returns \texttt{null} and the outer iteration is skipped without a full evaluation.
The two batch-level filters address distinct failure modes: the success constraint prevents regression; the recall gate prevents already-infeasible candidates from wasting the budget in global evaluation, keeping the finite inference budget focused on still-feasible improvements.
The complete procedure is described as \textsc{TrainBatch} in the supplementary material.

These two batch-level filters, together with the global level below, form REFINE's \textbf{constraint-decoupled architecture}, which fundamentally separates it from GEPA and ACE: the latter search for a single scalar objective in which recall is at most one tradable term, whereas REFINE treats $\text{recall} = 1.0$ as an inviolable hard constraint threaded through sampling, gating, and output.
A seemingly simpler alternative, i.e., folding the recall constraint into a scalar objective (a penalty term or constraint-dominance ranking) and then running evolutionary search, does not hold in text space: a single text edit often makes recall jump discretely rather than change smoothly, and this discrete jump makes penalty weights impossible to calibrate and causes dominance ranking to prematurely discard near-feasible, high-accuracy intermediate states.
REFINE therefore does not merge the objectives; instead, the Pareto elite pool preserves both feasible and near-feasible strategies, and the hard-constraint gate guarantees strict feasibility at output.

A skill returned by the gate is then evaluated on the complete $\mathcal{D}_{\text{evo}}$ and enters the \textbf{global elite pool} $\mathcal{P}$, which adopts GEPA's non-dominated retention strategy to maintain intermediate-state diversity and avoid improvement paths from being permanently closed.
The next outer iteration draws its parent from $\mathcal{P}$ by roulette (proportional to a cost-sensitive weighted score), so the outer loop explores across lineages while the inner loop deepens one; to prevent the search from collapsing onto a single high-scoring lineage, a per-parent selection cap $\kappa$ is enforced. Scoring weights and the formal Pareto definition are in the supplementary material.

After the entire evolution loop ends, the \textbf{global-level recall gate}, a direct extension of the batch-level gate, filters candidates in $\mathcal{P}$ satisfying $\mathrm{Recall}(S, \mathcal{D}_{\text{evo}})=1.0$ and outputs the one with the highest accuracy:
\begin{equation}
    S_{\text{best}} = \operatorname{\arg\max}_{S \in \mathcal{P},\, \mathrm{Recall}(S,\mathcal{D}_{\text{evo}})=1.0} \mathrm{Accuracy}(S).
\end{equation}

If no skill satisfies $\text{recall} = 1.0$ within the budget, the gate returns no skill, and the system falls back to routing alerts to human triage rather than deploying an unsafe skill that silently drops threats, i.e., the cost of falling back is reduced automation, never an undetected intrusion.
This \emph{fail-safe} direction is deliberate; generalization to unseen production traffic is bounded by post-deployment monitoring and rolling evolution (continuous alignment, below), not by the gate alone.

\paragraph{Continuous Alignment}
\label{sec:traj}

The operational scale is not static.
Risk standards shift with the threat landscape, asset priorities migrate with business changes, and threat-intelligence decay invalidates historical boundary samples.
An IOC marked as a threat requiring escalation in historical data becomes, once blocked at the perimeter, a dismissible false positive in today's alerts.
Accumulating historical boundary samples as ``anti-regression anchors'' at each re-evolution would lock the skill to an outdated operational scale.
REFINE therefore adopts \emph{re-sampling} rather than accumulation (in contrast to ACE-style rule accumulation) as its basic posture for continuous alignment: each re-evolution re-constructs $\mathcal{D}_{\text{evo}}$ from the latest time window so that the skill aligns with the current operational scale rather than a historical one.
GEPA and ACE stop at one-shot offline optimization with no post-deployment re-alignment; this continuous-alignment loop is a deployment-side capability unique to REFINE.

Re-evolution triggers fall into two complementary classes:
\begin{itemize}
    \item \textbf{Safety trigger (hard):} Sampled checks over alerts that the agent has auto-closed (taken over); any missed recall found therein means the recall hard constraint has been broken, and a new round is triggered immediately. This is post-hoc detection after the safety line has been crossed.
    \item \textbf{Timeliness trigger (soft):} A statistically significant drop in online accuracy indicates that the gap between the skill and the current operational scale has accumulated to the point of affecting overall triage quality; even before any confirmed missed recall, the skill is already outdated, and re-evolution should be triggered early to re-align before further degradation reaches the safety trigger.
\end{itemize}

After either trigger, REFINE re-samples $\mathcal{D}_{\text{evo}}$ from the latest time window of production disposal records and re-runs the global initial-skill generation together with the evolution loop on the new $\mathcal{D}_{\text{evo}}$, producing a new skill aligned with the current scale.
This process rolls continuously so that the skill evolves with the organization's operational standards rather than solidifying on a historical snapshot.
Deployment re-evolution triggers and the complete algorithm are given in the supplementary material.

\section{Experiments}

\subsection{Experimental Setup}

\paragraph{Datasets}
All the alert data and labels originate from a large enterprise's real-world online SOC; the labels are the analysts' daily disposal conclusions (dismiss/escalate), reflecting enterprise-specific asset boundaries and risk tolerance rather than a universal ground truth.
We build four alert scenarios spanning four MITRE ATT\&CK phases (Table~\ref{tab:scenarios}), each with a \emph{temporal split}: the evolution set $\mathcal{D}_{\text{evo}}$ ($n=100$) from an earlier window and the test set $\mathcal{D}_{\text{test}}$ ($n=150$) from a strictly later, non-overlapping one, mirroring REFINE's deployment (``evolve from past alerts, triage future ones'') and eliminating temporal leakage by construction.
Brief descriptions of the four scenarios are provided in the supplementary material.
Neither set is class-balanced, so the threat/benign ratios in Table~\ref{tab:scenarios} reflect each window's actual composition, from balanced (Phishing) to highly imbalanced (PROC2\_HEYE).
The $\text{recall}=1.0$ hard constraint is enforced only on $\mathcal{D}_{\text{evo}}$; $\mathcal{D}_{\text{test}}$ recall corroborates cross-time generalization.

\begin{table}[t]
    \centering
    \caption{Alert scenarios, ATT\&CK phases, and real class distributions (threat/benign).}
    \label{tab:scenarios}
    \small
    \setlength{\tabcolsep}{4pt}
    \begin{tabular}{l l c c}
        \toprule
        Scenario & ATT\&CK Phase & Evo (T/B) & Test (T/B) \\
        \midrule
        Phishing       & Initial Access    & 50 / 50 & 66 / 84  \\
        Host-Proc.     & Execution         & 40 / 60 & 60 / 90  \\
        NTLM-Hash      & Credential Access & 8 / 92  & 19 / 131 \\
        PROC2\_HEYE    & C\&C              & 9 / 91  & 5 / 145  \\
        \bottomrule
    \end{tabular}
\end{table}

\paragraph{Evaluation Metrics}
Recall is the fraction of true threats identified as escalated, $TP/(TP+FN)$; $\text{recall}=1.0$ is REFINE's hard constraint and the primary safety metric.
Accuracy is defined as $(TP+TN)/(TP+TN+FP+FN)$.
Under the $\text{recall}=1.0$ hard constraint, TP is saturated, so accuracy improvements come from correctly dismissed benign alerts (TN); higher accuracy thus directly reflects the ability to \emph{automatically close false positives}.

\paragraph{Baselines}
\textbf{(1) LLM Direct}: agent with no Skill; lower bound without operational calibration.
\textbf{(2) GEPA}~\cite{agrawal2026gepa}: A recent, representative self-evolution prompt optimization method using a Genetic-Pareto mechanism with reflective iteration; the same LLM, evolution set, and inference budget as REFINE, but without REFINE's global initial analysis, intelligent sampling, recall gate, or multi-round reflection.
\textbf{(3) ACE}~\cite{zhang2025agentic}: A self-evolution method that treats the in-context policy as an incrementally accumulated rule set, representing a strong baseline beyond prompt optimization; the same LLM, evolution set, and inference budget as REFINE, but without a recall hard constraint or gate.
\textbf{(4) REFINE (ours)}: Final Skill $S_{\text{best}}$ output by the full REFINE pipeline through the recall gate.

\paragraph{Implementation Details}
We use DeepSeek-V3.2 (671B MoE) as the agent's reasoning engine.
All self-evolution methods are given the same budget of 1000 single-sample inferences per scenario, which is the only explicit stopping condition, so the number of evolution iterations $N$ follows from the budget rather than being set by hand; evolution also stops early once evolution-set accuracy reaches 100\%.
The remaining parameters are batch size $|B|=5$, within-batch reflection rounds $R=3$, and per-parent selection cap $\kappa=2$, with roulette parent selection and intelligent sampling.
GEPA and ACE use the same LLM and the same inference budget; hyperparameters follow each method's recommended settings.
Crucially, neither GEPA nor ACE encodes recall as a hard constraint.
Both are driven by accuracy, a soft objective in which recall can be traded against precision.
Thus, the comparison isolates the effect of the recall hard constraint and REFINE's SOC-oriented sampling/reflection mechanisms rather than the underlying optimizer.
To eliminate the randomness of a single evolution run, all self-evolution methods (GEPA, ACE, REFINE) are run independently three times, and the main results (Table~\ref{tab:main}) report the mean over the three runs.

\begin{table*}[t]
    \centering
    \caption{Performance comparison (mean over three runs). (evo) = on $\mathcal{D}_{\text{evo}}$; (test) = on $\mathcal{D}_{\text{test}}$. Best results in each scenario in bold.}
    \label{tab:main}
    \small
    \begin{tabular}{l l c c c c}
        \toprule
        Scenario & Configuration & Accuracy (evo) & Recall (evo) & Accuracy (test) & Recall (test) \\
        \midrule
        Phishing  & LLM Direct                  & 0.550 & 0.987 & 0.490 & 0.977 \\
          & GEPA                        & 0.767 & 0.867 & \textbf{0.727} & 0.847 \\
          & ACE                         & \textbf{0.893} & 0.927 & 0.717 & 0.843 \\
          & Ours               & 0.780 & \textbf{1.000} & 0.630 & \textbf{1.000} \\
        \midrule
        Host-Proc.& LLM Direct                  & 0.660 & 0.813 & 0.720 & 0.773 \\
          & GEPA                        & 0.843 & 0.803 & 0.730 & 0.490 \\
          & ACE                         & \textbf{0.987} & 0.992 & 0.753 & 0.580 \\
          & Ours               & 0.763 & \textbf{1.000} & \textbf{0.777} & \textbf{0.807} \\
        \midrule
        NTLM-Hash & LLM Direct                  & 0.217 & 0.920 & 0.547 & 0.827 \\
          & GEPA                        & 0.920 & 0.793 & \textbf{0.873} & 0.490 \\
          & ACE                         & \textbf{0.993} & 0.958 & 0.817 & 0.917 \\
          & Ours               & 0.907 & \textbf{1.000} & 0.537 & \textbf{1.000} \\
        \midrule
        PROC2\_HEYE & LLM Direct                & 0.917 & 0.890 & 0.887 & \textbf{1.000} \\
          & GEPA                        & 0.963 & 0.813 & 0.987 & 0.933 \\
          & ACE                         & \textbf{1.000} & \textbf{1.000} & \textbf{0.990} & \textbf{1.000} \\
          & Ours               & 0.990 & \textbf{1.000} & 0.983 & \textbf{1.000} \\
        \bottomrule
    \end{tabular}
\end{table*}

\subsection{Main Results}

\paragraph{Evolution Effectiveness}

Table~\ref{tab:main} reports four configurations across all four scenarios (mean of three independent runs).
Only REFINE consistently satisfies the zero-missed-recall constraint on the evolution set, and it is the most accurate strategy within the feasible region: only REFINE reaches $\text{recall}=1.0$ in all four scenarios, while the other three configurations each miss threats in at least one.
ACE tops evolution-set accuracy on Phishing, Host-Process, and NTLM-Hash (0.893 / 0.987 / 0.993), but two of those three have evolution-set recall below 1.0 (0.927 / 0.958), placing its higher accuracy outside the feasible region.
Among strategies that satisfy $\text{recall}=1.0$, REFINE attains the highest evolution-set accuracy: NTLM-Hash is emblematic, where it raises accuracy from 0.217 (LLM Direct) to 0.907 while holding $\text{recall}=1.0$, turning an almost ``escalate-all'' scenario into one with substantial false-positive takeover space.

Cross-time behavior reflects the twofold value of the recall hard constraint---stable generalization and controllable failure direction.
On the strictly later test window, REFINE maintains $\text{recall}=1.0$ in Phishing, NTLM-Hash, and PROC2\_HEYE; in Host-Process, test recall is 0.807, the highest among all methods in that scenario.
Hard scenarios still exhibit overfitting (NTLM-Hash drops from evo 0.907 to test 0.537; see the NTLM-Hash case study for attribution), but the direction of degradation differs sharply: REFINE loses only in accuracy and conservatively (more benign alerts escalated) while holding recall at 1.0, whereas ACE's degradation erodes recall (Host-Process 0.992$\to$0.580) in the direction of missing threats---the same overfitting with very different safety consequences.
This addresses overfitting concerns and shows that zero missed recall on the evolution set cannot formally guarantee it on future windows; Skills must keep re-evolving via continuous alignment.

\begin{table*}[t]
    \centering
    \caption{Selection-fairness check against the strongest baseline (ACE). We apply REFINE's evo $\text{recall}=1.0$ gate to ACE's own pool, take the most accurate feasible candidate, and report the same performance columns as Table~\ref{tab:main}; $\emptyset$ = no feasible candidate, and ACE means are over its feasible runs only. Feasibility rate (as in Table~\ref{tab:ablation}) is 0.000 / 0.667 / 0.667 / 1.000 for ACE across the four scenarios and 1.000 throughout for REFINE. Bold = better test recall (the safety metric).}
    \label{tab:fairness}
    \small
    \begin{tabular}{l l c c c c}
        \toprule
        Scenario & Configuration & Accuracy (evo) & Recall (evo) & Accuracy (test) & Recall (test) \\
        \midrule
        Phishing     & ACE (gated)     & $\emptyset$ & $\emptyset$ & $\emptyset$ & $\emptyset$ \\
             & Ours & 0.780 & 1.000 & 0.630 & \textbf{1.000} \\
        \midrule
        Host-Process & ACE (gated)     & 0.990 & 1.000 & 0.760 & 0.575 \\
             & Ours & 0.763 & 1.000 & 0.777 & \textbf{0.807} \\
        \midrule
        NTLM-Hash    & ACE (gated)     & 0.995 & 1.000 & 0.790 & 0.921 \\
             & Ours & 0.907 & 1.000 & 0.537 & \textbf{1.000} \\
        \midrule
        PROC2\_HEYE  & ACE (gated)     & 1.000 & 1.000 & 0.985 & 1.000 \\
             & Ours & 0.990 & 1.000 & 0.983 & 1.000 \\
        \bottomrule
    \end{tabular}
\end{table*}

\subsection{Comparison within Feasible Region}
\label{subsec:fairness}

REFINE's output in Table~\ref{tab:main} is chosen by the recall gate, whereas ACE reports its accuracy-best output. This invites a challenge: is REFINE's recall advantage merely a selection artifact, with a $\text{recall}=1.0$ Skill already sitting unselected in ACE's pool? To test this, we apply REFINE's gate to ACE's own search pool (evolution-set $\text{recall}=1.0$, most accurate candidate) and report that candidate's out-of-sample test behavior (Table~\ref{tab:fairness}); selection uses evolution-set metrics only, with no test-set leakage.

Post-hoc selecting ACE's feasible candidate still does not recover REFINE's generalizable recall safety: in Phishing, ACE's pool contains no feasible candidate at all ($\emptyset$, feasibility rate 0.000); where ACE does find one (Host-Process, NTLM-Hash) it is overfit, its test recall dropping to $0.575$ and $0.921$ versus REFINE's $0.807$ and $1.000$; only in the easiest scenario (PROC2\_HEYE) do the two tie. The advantage is therefore not a selection artifact: because ACE never explicitly enforces the feasible region, even candidates that land inside it fail to keep recall across time, whereas REFINE threads the gate through sampling, batch, and global search to yield a Skill that is both feasible and recall-generalizing.

\begin{table*}[t]
    \centering
    \caption{Ablation results on the Host-Process scenario. All values except feasibility rate are means over three runs.}
    \label{tab:ablation}
    \small
    \begin{tabular}{lccccc}
        \toprule
        Configuration & Iteration Success & Waste & Accuracy (evo) & Recall (evo) & Feasibility \\
        \midrule
        Full & 1.000 & 0.000 & 0.763 & 1.000 & 1.000 \\
        w/o initial skill & 0.958 & 0.000 & 0.783 & 0.933 & 0.000 \\
        w/o intelligent sampling & 0.660 & 0.000 & 0.807 & 0.933 & 0.000 \\
        w/o initial skill \& intelligent sampling & 0.623 & 0.000 & 0.803 & 0.817 & 0.000 \\
        w/o multi-ref \& recall\_gate & 0.963 & 0.245 & 0.790 & 0.975 & 0.333 \\
        w/o all & 0.377 & 0.000 & 0.693 & 0.808 & 0.000 \\
        \bottomrule
    \end{tabular}
\end{table*}

\subsection{Ablation Study}

We conduct ablations on the Host-Process scenario, which depends on process chains, command-line intent, and environmental context, and therefore stresses both the recall safety baseline and benign-alert dismissal.
Except for the removed mechanisms, all configurations use the same data split, inference budget, Pareto elite pool, and per-parent selection cap $\kappa=2$, running independently three times; because the budget rather than a fixed iteration count is held constant, arms that waste budget on infeasible candidates complete fewer iterations.
The specific ablation arms and the three diagnostic metrics (\textbf{iteration success rate}, \textbf{wasted rate}, \textbf{feasibility rate}) are defined in the supplementary material.

Table~\ref{tab:ablation} provides component-level evidence for the full design.

\textbf{Global initial Skill.}
Removing the initial Skill preserves a relatively high iteration-success rate ($0.958$) and even raises accuracy to $0.783$, but recall falls to $0.933$ and no run produces a feasible output.
The global initialization therefore does more than start the search: it supplies a full-distribution prior that keeps local reflections anchored to the recall-safe region.

\textbf{Intelligent sampling.}
Removing intelligent sampling sharply reduces iteration success from $1.000$ to $0.660$; recall again reaches only $0.933$, with zero feasibility.
When both the initial Skill and intelligent sampling are absent, iteration success declines further to $0.623$ and recall to $0.817$.
This pattern shows that intelligent sampling directs reflection toward the missed-recall and false-positive boundaries that must be repaired, while the initial Skill supplies the global context in which those local repairs remain valid.

\textbf{Multi-round reflection and recall gate.}
Removing this pair lowers iteration success from $1.000$ to $0.963$, reduces recall to $0.975$, yields feasible outputs in only one of three runs, and is the only setting with nonzero waste ($0.245$).
These diagnostics match the intended division of labor: multi-round reflection raises the chance that local revision succeeds, whereas the recall gate rejects infeasible candidates before they consume further evaluation budget.
Because the two mechanisms are ablated together, this experiment supports that joint explanation but does not separately quantify either mechanism's effect; that would require individual \emph{w/o multi-ref} and \emph{w/o recall gate} arms.

\textbf{Overall interaction under the hard constraint.}
Removing all mechanisms collapses iteration success to $0.377$, recall to $0.808$, and feasibility to $0$.
Across every ablation, accuracy can equal or exceed full REFINE (up to $0.807$ versus $0.763$) while recall remains below $1.0$.
The evidence therefore supports the intended division of labor: the initial Skill, intelligent sampling, and the multi-reflection--gate pair contribute distinct search guidance or constraint control, and their combination is required to reliably return a deployable recall-safe Skill within budget.

\subsection{Model Robustness}
\label{subsec:robustness}

To test whether REFINE's gains depend on a single reasoning engine, we replace DeepSeek-V3.2 with Kimi-K2.6, an independently trained open-weight MoE model from a different vendor, and re-run REFINE and the LLM Direct lower bound on all four scenarios (Table~\ref{tab:robustness}, evolution set, mean over three runs).

\begin{table}[t]
    \centering
    \caption{Model robustness on Kimi-K2.6 (evolution set, mean over three runs). Best per column within each scenario in bold.}
    \label{tab:robustness}
    \small
    \setlength{\tabcolsep}{4pt}
    \begin{tabular}{l l c c}
        \toprule
        Scenario & Configuration & Accuracy (evo) & Recall (evo) \\
        \midrule
        Phishing     & LLM Direct    & 0.730 & \textbf{1.000} \\
             & Ours & \textbf{0.960} & \textbf{1.000} \\
        \midrule
        Host-Proc.   & LLM Direct    & 0.790 & 0.900 \\
             & Ours & \textbf{0.840} & \textbf{1.000} \\
        \midrule
        NTLM-Hash    & LLM Direct    & 0.220 & 0.708 \\
             & Ours & \textbf{0.983} & \textbf{1.000} \\
        \midrule
        PROC2\_HEYE  & LLM Direct    & 0.930 & 0.670 \\
             & Ours & \textbf{1.000} & \textbf{1.000} \\
        \bottomrule
    \end{tabular}
\end{table}

On Kimi-K2.6, REFINE again reaches $\text{recall}=1.0$ on all four evolution sets (3/3 runs each), matching its DeepSeek-V3.2 behavior, whereas the uncalibrated LLM Direct misses recall in three of the four scenarios (Host-Process $0.900$, NTLM-Hash $0.708$, PROC2\_HEYE $0.670$).
Within the $\text{recall}=1.0$ feasible region, REFINE substantially expands benign false-positive auto-closure, most strikingly on NTLM-Hash, where evolution-set accuracy rises from $0.220$ to $0.983$, echoing the $0.217\rightarrow0.907$ pattern on DeepSeek-V3.2.
This indicates that the ability to reach the constrained optimum is not specific to one reasoning engine.

Because DeepSeek-V3.2 and Kimi-K2.6 are independently trained MoE models from different vendors, consistently attaining the recall hard constraint on both suggests the gains stem from REFINE's mechanism rather than a model-specific artifact.
Privacy rules out closed APIs (SOC alert data cannot leave the enterprise), so we evaluate two representative self-hostable open-weight engines; broader capability-tier coverage is a natural next step.
Consistent with the rest of the paper, the robustness claim is anchored on the evolution set, with cross-time generalization analyzed on DeepSeek-V3.2 in the main-results and case-study sections.

\subsection{Effect of Skill and Context Size}

REFINE and ACE both improve agent behavior through reflective iteration, but the carrier of improvement differs fundamentally: ACE treats the policy as an ever-accumulating rule set (a playbook), whereas REFINE maintains a compact, structured, partitioned Skill.
Table~\ref{tab:size} compares the token size of each method's final artifact per scenario.

\begin{table}[t]
    \centering
    \caption{Final-artifact token size (cl100k\_base \cite{openai_tiktoken}, mean over three runs). The artifact is ACE's accumulated playbook and REFINE's output skill $S_{\text{best}}$.}
    \label{tab:size}
    \small
    \setlength{\tabcolsep}{2.5pt}
    \begin{tabular}{l c c c}
        \toprule
        Scenario & ACE & Ours & ACE/Ours \\
        \midrule
        Phishing     & 58{,}070 & 4{,}707 & 12.3$\times$ \\
        Host-Process & 42{,}147 & 7{,}586 & 5.6$\times$ \\
        NTLM-Hash    & 24{,}249 & 4{,}108 & 5.9$\times$ \\
        PROC2\_HEYE  & 18{,}026 & 6{,}223 & 2.9$\times$ \\
        \bottomrule
    \end{tabular}
\end{table}

ACE's playbook (18K--58K tokens) expands monotonically with scenario difficulty, whereas REFINE's Skill stays within 4.1K--7.6K tokens, a 2.9$\times$--12.3$\times$ gap. The gap reflects \emph{bloat and overfitting sharing a source} (ACE accumulates rules to fit the evolution distribution, but the bloated set fails to transfer, degrading test accuracy and recall) and \emph{deployment cost}: a playbook fed into inference with every alert inflates continual overhead and is hard to audit, whereas a few-thousand-token structured Skill can be read and revised directly by operators.

\section{Conclusion}

We presented REFINE, which continuously evolves structured Skills to align LLM Agent alert triage with enterprise operational scales.
REFINE localizes judgment blind spots from the global alert distribution and current error boundary, and constrains evolution through a recall gate.
Experiments across four alert scenarios in a real industrial SOC show that REFINE maintains $\text{recall}=1.0$ on every evolution set while improving benign-alert closure; results on later time windows further provide evidence of cross-time transfer.

\bibliography{refs}

\clearpage
\appendix

\section{Key Notation}
\label{sec:notation}

\begin{table}[ht]
\centering
\caption{Key notation used throughout the paper.}
\small
\begin{tabular}{l p{6cm}}
\toprule
\textbf{Symbol} & \textbf{Description} \\
\midrule
$x$ & An alert sample \\
$y \in \{0, 1\}$ & Ground-truth label; $1$ = escalate, $0$ = dismiss \\
$\hat{y}$ & Agent-predicted triage label \\
$S$ & A Skill (structured natural-language Agent behavior specification encoding judgment rules, evidence-chain reasoning, and tool invocation protocols) \\
$S_0$ & Initial Skill generated via global analysis \\
$S_{\text{best}}$ & Final output Skill satisfying recall hard constraint \\
$\mathcal{D}_{\text{evo}}$ & Evolution set  \\
$\mathcal{D}_{\text{test}}$ & Held-out test set, used only for one-shot final evaluation; results reported and never fed back \\
$B$ & A batch of samples drawn from $\mathcal{D}_{\text{evo}}$ \\
$\mathcal{P}$ & Global Pareto elite pool \\
$f_S(x)$ & Agent's predicted label for $x$ under Skill $S$ \\
$N$ & Maximum number of evolution iterations \\
$R$ & Maximum reflection rounds per batch \\
$\kappa$ & Per-parent selection cap; a node selected as parent $\kappa$ times is excluded from further parent selection \\
$\mathcal{H}$ & Evolution trajectory (per-iteration records of batch, Skill, and metrics) \\
\bottomrule
\end{tabular}
\end{table}

\section{Threat Model and Assumptions}
\label{sec:threat-model}

REFINE operates within the following environmental assumptions and trust boundaries.

\paragraph{Deployment Scope}
REFINE targets enterprise-internal SOC triage workflows.
The evolution set and test set are drawn from the same alert type (one rule or detection scenario per deployment), and both sets are sampled from non-overlapping time windows to prevent temporal leakage.

\paragraph{Label Source and Quality}
Labels are derived from analyst operational disposals (escalate/dismiss) rather than specially annotated ground truth.
They reflect the operational judgment scale of the organization---asset priorities, risk tolerance, and disposal conventions---rather than universal security standards.
Label inconsistency is treated as \emph{operational judgment-scale divergence} rather than annotation noise: the same alert may legitimately receive different labels from different analysts due to differing contexts or standards.
REFINE does not require labels to be globally noise-free across operators or time; it performs constrained evolution against the operational labels available in each evolution window.

\paragraph{Adversarial Assumptions}
REFINE assumes no adversarial manipulation of the evolution set or labels.
Alert data is assumed to be generated by the production detection system and reflects genuine operational traffic.
Adversarial evasion of the deployed Skill (e.g., crafting alerts to exploit Skill blind spots) is outside the scope of this work.

\paragraph{Tool Availability}
The Agent may invoke a fixed set of read-only security tools (threat intelligence, asset lookup, file hash queries, web search).
Tool outputs are assumed to be accurate and available; tool failure or adversarial tool responses are not modeled.

\section{Batch Training Procedure}
\label{sec:alg-trainbatch}

\begin{algorithm}[ht]
\caption{\textsc{TrainBatch}}
\label{alg:trainbatch}
\small
\begin{algorithmic}[1]
\REQUIRE Batch $B$, parent Skill $S_{\mathrm{parent}}$, max rounds $R$
\ENSURE Best recall-safe Skill for this batch, or \texttt{null}
\STATE $S_{\mathrm{cur}} \leftarrow S_{\mathrm{parent}}$; $\mathcal{C} \leftarrow \emptyset$ \COMMENT{Skills produced in this batch}
\STATE $a^{\star} \leftarrow \mathrm{Accuracy}(S_{\mathrm{parent}}, B)$; $r^{\star} \leftarrow \mathrm{Recall}(S_{\mathrm{parent}}, B)$ \COMMENT{Batch-level best so far}
\FOR{$r = 1$ \TO $R$}
    \STATE Retrieve cached Agent reasoning trajectories of $S_{\mathrm{cur}}$ on $B$
    \STATE $S' \leftarrow \text{Reflect}(B, S_{\mathrm{cur}}, \text{trajectories})$
    \STATE Evaluate $S'$ on $B$; compute $\mathrm{Accuracy}(S', B)$ and $\mathrm{Recall}(S', B)$
    \IF{$\mathrm{Recall}(S', B) \ge r^{\star}$ \AND ($\mathrm{Accuracy}(S', B) > a^{\star}$ \OR $\mathrm{Recall}(S', B) > r^{\star}$)}
        \STATE \COMMENT{\textbf{Success constraint}: recall does not drop and $\ge 1$ metric strictly improves}
        \STATE $\mathcal{C} \leftarrow \mathcal{C} \cup \{S'\}$; $S_{\mathrm{cur}} \leftarrow S'$ \COMMENT{Extend the same lineage in depth}
        \STATE $a^{\star} \leftarrow \max(a^{\star}, \mathrm{Accuracy}(S', B))$; $r^{\star} \leftarrow \mathrm{Recall}(S', B)$
    \ELSE
        \STATE \textbf{break} \COMMENT{Within-batch regression; stop this batch}
    \ENDIF
\ENDFOR
\STATE $S^* \leftarrow \arg\max_{S \in \mathcal{C},\, \mathrm{Recall}(S, B) = 1.0} \mathrm{Accuracy}(S, B)$ \COMMENT{Batch-level recall gate}
\IF{$S^*$ exists}
    \RETURN $S^*$
\ELSE
    \RETURN \texttt{null} \COMMENT{Recall gate not met}
\ENDIF
\end{algorithmic}
\end{algorithm}

\section{Deployment Re-Evolution Triggers}
\label{sec:closed-loop}

REFINE's deployment loop comprises two re-evolution trigger paths.
Both triggers are grounded in observable online performance signals.

\paragraph{Trigger Path 1: Safety Trigger (Hard)}
After the skill is deployed, the agent auto-closes (takes over) a portion of online alerts.
Sampled checks over these ``taken-over'' alerts: any missed recall found therein---an alert that should have been escalated but was auto-closed---means the recall hard constraint has been broken, and a new evolution round is triggered immediately.
The trigger then re-samples $\mathcal{D}_{\text{evo}}$ from the latest time window of production disposal records and re-runs the global generation and evolution loop on the new $\mathcal{D}_{\text{evo}}$.

\paragraph{Trigger Path 2: Timeliness Trigger (Soft)}
In production, REFINE monitors online accuracy (or takeover rate).
A statistically significant drop in accuracy (e.g., exceeding two standard deviations from the rolling baseline) indicates that the gap between the Skill and the current operational scale has accumulated to the point of affecting overall triage quality---even before any confirmed missed recall, the Skill is already outdated, and re-evolution is triggered early to re-align with the current scale before further degradation reaches the safety trigger.

\section{The REFINE Algorithmic Procedure}
\label{sec:alg-refine}

\begin{algorithm}[ht]
\caption{REFINE}
\label{alg:refine}
\small
\begin{algorithmic}[1]
\REQUIRE Evolution set $\mathcal{D}_{\text{evo}}$, iterations $N$, reflection rounds $R$, per-parent selection cap $\kappa$
\ENSURE Recall-constrained best Skill $S_{\text{best}}$, evolution trajectory $\mathcal{H}$
\STATE \textbf{// Phase 1: Global initial Skill generation}
\STATE $S_0 \leftarrow \text{GenerateInitialSkill}(\mathcal{D}_{\text{evo}})$
\STATE \textbf{// Phase 2: Initialize elite pool and trajectory}
\STATE $\mathcal{P} \leftarrow \{S_0\}$; $\mathcal{H} \leftarrow \emptyset$
\STATE $\textit{metrics}_0 \leftarrow \text{Evaluate}(S_0, \mathcal{D}_{\text{evo}})$
\STATE \textbf{// Phase 3: Self-evolution iterations}
\FOR{$i = 1$ \TO $N$}
    \STATE $S_{\mathrm{parent}} \leftarrow \text{SelectParent}(\mathcal{P}, \kappa)$ \COMMENT{Roulette over $\mathcal{P}$; parents already selected $\kappa$ times are excluded, forcing exploration}
    \STATE $B \leftarrow \text{SmartSample}(\mathcal{D}_{\text{evo}}, S_{\mathrm{parent}})$ \COMMENT{Priority P1 $>$ P2 $>$ P3}
    \STATE $S_{\mathrm{batch}} \leftarrow \text{TrainBatch}(B, S_{\mathrm{parent}}, R)$
    \IF{$S_{\mathrm{batch}} = \texttt{null}$}
        \STATE \textbf{continue} \COMMENT{Recall gate not met}
    \ENDIF
    \STATE $\textit{metrics} \leftarrow \text{Evaluate}(S_{\mathrm{batch}}, \mathcal{D}_{\text{evo}})$
    \STATE $\mathcal{H} \leftarrow \mathcal{H} \cup \{\text{RecordIteration}(i, B, S_{\mathrm{batch}}, \textit{metrics})\}$
    \STATE $\mathcal{P} \leftarrow \text{UpdateElitePool}(\mathcal{P}, S_{\mathrm{batch}}, \textit{metrics})$
    \IF{$\mathrm{Accuracy}(S_{\mathrm{batch}}, \mathcal{D}_{\text{evo}}) = 1.0$}
        \STATE \textbf{break}
    \ENDIF
\ENDFOR
\STATE \textbf{// Phase 4: Output}
\STATE $S_{\text{best}} \leftarrow \arg\max_{S \in \mathcal{P},\, \mathrm{Recall}(S, \mathcal{D}_{\text{evo}}) = 1.0} \mathrm{Accuracy}(S)$
\IF{$S_{\text{best}}$ exists}
    \RETURN $S_{\text{best}}, \mathcal{H}$
\ELSE
    \RETURN \texttt{null}, $\mathcal{H}$ \COMMENT{No feasible Skill within budget; fail-safe fallback to human triage}
\ENDIF
\end{algorithmic}
\end{algorithm}

\section{Dataset Statistics and Split Rationale}
\label{sec:datasets}

\begin{table}[ht]
\centering
\caption{Dataset class distributions under temporal split. Esc.\ = Escalate, Dis.\ = Dismiss. $\mathcal{D}_{\text{evo}}$ is the earlier window and $\mathcal{D}_{\text{test}}$ the strictly later window; both preferentially sample the real threats available in their own window and are not class-rebalanced, with $\mathcal{D}_{\text{evo}}$ additionally augmented with historical threats from within its window.}
\label{tab:datasets}
\small
\begin{tabular}{l l r r r r}
\toprule
\textbf{Scenario} & \textbf{Split} & \textbf{$n$} & \textbf{Esc.} & \textbf{Dis.} & \textbf{Esc.\%} \\
\midrule
Phishing  & $\mathcal{D}_{\text{evo}}$  & 100 & 50  & 50  & 50.0\% \\
          & $\mathcal{D}_{\text{test}}$ & 150 & 66  & 84  & 44.0\% \\
\midrule
Host-Process & $\mathcal{D}_{\text{evo}}$  & 100 & 40  & 60  & 40.0\% \\
          & $\mathcal{D}_{\text{test}}$ & 150 & 60  & 90  & 40.0\% \\
\midrule
NTLM-Hash & $\mathcal{D}_{\text{evo}}$  & 100 & 8   & 92  & 8.0\%  \\
          & $\mathcal{D}_{\text{test}}$ & 150 & 19  & 131 & 12.7\% \\
\midrule
PROC2\_HEYE & $\mathcal{D}_{\text{evo}}$  & 100 & 9   & 91  & 9.0\%  \\
          & $\mathcal{D}_{\text{test}}$ & 150 & 5   & 145 & 3.3\%  \\
\bottomrule
\end{tabular}
\end{table}

\paragraph{Temporal Split Protocol}
For each scenario, alerts are ordered by their generation timestamp and cut at a fixed split point: all alerts before the cut form the evolution window, all alerts after it form the test window. From these two disjoint windows we draw fixed-size samples---$\mathcal{D}_{\text{evo}}$ ($n{=}100$) from the earlier window and $\mathcal{D}_{\text{test}}$ ($n{=}150$) from the strictly later window---so the two sets are disjoint in time. The sizes are fixed across all four scenarios to keep evaluation budget and cross-scenario comparability consistent; the temporal split (which time window) and the fixed sample size (how many) are orthogonal choices. This mirrors the production workflow (evolve on historical alerts, deploy on future alerts) and rules out temporal leakage: no information from the test window can enter evolution.

\paragraph{Within-Window Threat Augmentation and Evo-only Aggregation}
Both $\mathcal{D}_{\text{evo}}$ and $\mathcal{D}_{\text{test}}$ preferentially sample the real threat (escalate) samples available in their respective windows and are never rebalanced to a fixed positive ratio; neither reflects the raw positive rate of the underlying SOC stream. The two sets differ in two preprocessing steps applied \emph{only} to $\mathcal{D}_{\text{evo}}$: (i) representative-aggregation, which removes redundancy and exposes diverse defects for evolution, and (ii) within-window threat augmentation (below). $\mathcal{D}_{\text{test}}$ receives neither; it simply takes the threats available in its later window.
$\mathcal{D}_{\text{evo}}$ is additionally augmented with \emph{real} historical threat (escalate) samples to obtain enough defect signal for reflection-driven evolution; these are distinct real alerts, not duplicated instances. The augmentation follows a single rule: positives are added \emph{up to at most 50\% of the evolution set size}; when real threats are insufficient to reach this cap, we use as many as exist. This is why the four scenarios differ in evo positive rate: Phishing reaches the 50\% cap, while Host-Process (40\%), PROC2\_HEYE (9\%), and NTLM-Hash (8\%) are limited by the real threats available in the window. Crucially, all added positives are drawn \emph{exclusively from the earlier evolution window}, never from the test window, so augmentation introduces no future leakage and no train--test overlap.

\begin{table*}[t]
\centering
\caption{NTLM-Hash evolution trajectory (per-node, evolution set).}
\label{tab:trajectory}
\footnotesize
\setlength{\tabcolsep}{4pt}
\begin{tabularx}{\textwidth}{c c r r >{\raggedright\arraybackslash}X}
\toprule
\textbf{Node} & \textbf{Parent} & \textbf{Acc} & \textbf{Rec} & \textbf{Key Skill change} \\
\midrule
0 & --- & 0.56 & 1.00 & Global init; escalates nearly all alerts (8 threats recalled, 44 benign wrongly escalated) \\
1 & 0 & 0.86 & 0.75 & Introduces ``strong trusted local feature chain'' exclusion (four conditions required); downgrades VirusTotal to auxiliary evidence. Accuracy leaps, but the trusted rule over-suppresses and recall regresses \\
2 & 1 & 0.22 & 0.875 & First expansion of node~1: repairs recall but draws the ``trusted'' boundary too tight; batch acc 1.0 yet global collapse to 0.22, Pareto-dominated by node~0 \\
3 & 1 & 0.21 & 1.00 & Second (and final, under cap $\kappa{=}2$) expansion of node~1: re-escalates nearly all alerts; batch acc 1.0 yet global collapse to 0.21, also dominated by node~0. Node~1 is now capped and excluded, so the search re-expands from node~0 \\
4 & 0 & 0.91 & 0.875 & Re-expands from node~0: ``behavior pattern over single indicators + complete risk-exclusion chain,'' translating stability-contract boundary~1 (unknown/non-mainstream software accessing lsass must be recalled) \\
5            & 4     & 0.94 & 0.875 & Refines trusted-identity assessment (valid signature + vendor standard path as core evidence, de-emphasizing companyName); \textbf{globally weighted-optimal}, but recall${}<1.0$ \\
6            & 5     & 0.87 & 1.00 & Repairs the identity-missing missed recall (non-mainstream software with valid signature but missing identity info must be recalled); recall restored, accuracy slightly lower \\
7 (final $S_{\text{best}}$) & 5 & 0.91 & 1.00 & Tightens the trusted-identity condition to fix the same missed recall; accuracy 0.91, recall$=1.0$; weighted-optimal, selected as $S_{\text{best}}$ through the recall gate \\
\bottomrule
\end{tabularx}
\end{table*}

\paragraph{Why is Recall Anchored on the Evolution Set?}
Test-window positive counts vary widely across scenarios (5--66), reflecting how many real threats each later window actually contained rather than any fixed target ratio; PROC2\_HEYE's later window, for instance, yielded only 5. Neither set is rebalanced to a fixed positive ratio.
Recall is anchored on $\mathcal{D}_{\text{evo}}$ for a structural reason rather than a statistical one: evolution can only observe the evolution window, whereas $\mathcal{D}_{\text{test}}$ is a strictly later, unseen future window held out to rule out temporal leakage. The recall$=1.0$ hard constraint is therefore enforced where optimization actually happens ($\mathcal{D}_{\text{evo}}$), and test-set recall serves as an out-of-time check that the evolved Skill still holds on future alerts.

\section{Dataset Scenario Descriptions}
\label{sec:scenario-desc}

The four scenarios were selected to cover distinct alert evidence patterns rather than to sample the SOC stream uniformly.

\paragraph{Phishing} Initial-access email alerts, where the main evidence comes from sender identity, subject/body phrasing, and link or credential-request cues. This scenario is the most text-like of the four and has the most balanced class split.

\paragraph{Host-Process} Endpoint execution alerts, where process ancestry, command line, parent-child relationships, and execution context dominate the decision. This scenario stresses reasoning over behavior chains rather than single indicators.

\paragraph{NTLM-Hash} Credential-access alerts centered on suspicious access to LSASS or related credential material. This scenario is the most brittle across time in the paper and is used as the main trajectory case study.

\paragraph{PROC2\_HEYE} Command-and-control alerts with the strongest class imbalance in the set. This scenario mainly tests whether the Skill can preserve recall when positive evidence is sparse and benign traffic dominates.

\section{NTLM-Hash Evolution Trajectory}
\label{sec:trajectory}

Table~\ref{tab:trajectory} gives the complete per-node trajectory of a single representative evolution run on NTLM-Hash over the evolution set ($\mathcal{D}_{\text{evo}}$, $n=100$). Nodes are numbered by the expansion order of the Pareto global search (node~0 corresponds to the initial Skill $S_0$); the ``Parent'' column reconstructs the search-tree structure (including the dead branch at nodes~2--3 and the backtracking to node~0).

Under Pareto global search, the trajectory follows a \emph{leap--regression--convergence} pattern. Node~1 raises accuracy from 0.56 to 0.86, but the ``strong trusted rule'' over-suppresses and recall regresses to 0.75; nodes~2--3 are the two children node~1 spawns before it reaches the per-parent selection cap ($\kappa{=}2$), each reaching batch accuracy 1.0 on its 5-sample batch yet collapsing to 0.21--0.22 on the full evolution-set evaluation. Two mechanisms then jointly steer the search back to $S_0$: node~1, though holding the highest weighted score, becomes ineligible as a parent once it hits the cap, and its two children are Pareto-dominated by $S_0$ (higher on neither accuracy nor recall), leaving $S_0$ as the sole elite among the uncapped candidates. This exposes nodes~2--3 as a dead-end branch and underscores the value of global evaluation over the batch-local view: a reflection that is self-consistent within its 5-sample batch need not survive full evolution-set evaluation. Re-expanded from $S_0$, nodes~4$\to$5 attain 0.91$\to$0.94 (weighted-optimal) with recall stable at 0.875 via ``behavior pattern first + risk-exclusion chain + contract-boundary translation''; the final reflection localizes and repairs the identity-missing missed recall (valid signature but missing identity information), and node~7 restores recall$=1.0$ while keeping accuracy 0.91. Notably, node~5 with the higher accuracy (0.94) is excluded by the recall gate because recall$=0.875$, and the final output is node~7 (0.91) with recall$=1.0$ as $S_{\text{best}}$---reflecting that the recall gate only filters and does not trade recall for accuracy.

\section{Attributing the NTLM-Hash Test Accuracy Drop}
\label{sec:case-study-ntlm}

REFINE's NTLM-Hash accuracy dropping from evolution-set 0.907 to test-set 0.537 is the largest cross-time degradation in the paper. Both windows are dominated by the same alert rule (95/100 evo, 137/150 test), ruling out a cross-rule shift; the actual drift is an \emph{intra-rule reshaping of benign trigger sources}---87.7\% of test-window benign alerts come from processes unseen during training (new third-party security agents and newly onboarded employee applications). The failure is strictly confined to this training-unseen distribution: the Skill correctly ignores 87.5\% of training-seen processes but only 47.4\% of unseen ones, and 95.2\% of its test FPs concentrate there (correcting them would raise test accuracy to 0.980). This is exactly the operational-scale drift of this framework, and empirically grounds the need for continual alignment.

\section{Necessity of Decoupling Scoring, Pareto Retention, and the Hard-Constraint Gate}
\label{sec:decoupling}

This appendix elaborates the two-failure-mode argument for the framework's batch/global search: why the two standard paths for collapsing multi-objective constrained optimization into a single solvable problem are ill-suited to text-space Skill optimization.

\subsection{Discrete-Jump Observation}

Skill optimization operates in natural-language text space. Unlike continuous parameter spaces, recall changes induced by text edits are discrete: a single text modification can drop recall from 1.0 to 0.75 in one step (NTLM-Hash $S_0 \to$ node~1), with no intermediate recall value along that edit. This observation is not a formal property of all text spaces---some Skill classes may permit fine-grained adjustment---but as long as indivisible boundary jumps exist, the two standard methods below fail structurally.

\subsection{Failure Mode 1: Penalty Scalarization}

The standard way to fold multi-objective (Acc, Recall) into a single objective is a penalty:
$$\max_S \mathrm{Acc}(S) - \lambda \cdot \max(0, 1-\mathrm{Recall}(S)).$$
This form admits no workable $\lambda$ in text space. Recall's discrete jumps provide no gradient signal to tune against: too-small $\lambda$ degenerates into unconstrained optimization (any Acc-for-Recall trade is accepted), while too-large $\lambda$ collapses onto the trivial feasible solution (escalate everything, minimizing Acc but achieving Recall$=$1.0). No middle regime exists: at a jump point, recall changes by 0.25 in one step, so the $\lambda$ threshold is a single point rather than an interval. In other words, the smooth cost-constraint tradeoff that penalty methods require does not hold in text space.

\subsection{Failure Mode 2: Constraint-Dominance Ranking}

The standard treatment of multi-objective + hard-constraint is constraint-dominance ranking: sort by constraint violation first, then by Pareto relation among equally-violating solutions. This method treats all infeasible solutions as equivalent---any recall${}<1.0$ is infeasible, with no distinction between recall$=0.875$ and recall$=0.75$.

This equivalence is harmless in continuous spaces (constraint violation is smoothly measurable, and near-feasible solutions are close to the feasible region), but in text space it discards key genetic information. A strategy with recall$=0.875$ is one rule away from feasibility; the ``almost-feasible'' information carried by that rule matters precisely because no smooth path from 0.875 back to 1.0 exists---the next edit either jumps back to 1.0 or drops to 0.75. Constraint-dominance ranking deletes these ``one-step-away'' strategies as equivalent to ``far-away'' ones, discarding the most valuable evolutionary intermediate states in text space.

\subsection{Derivation of the Decoupled Architecture}

The two failure modes above jointly point to a design direction: decoupling \emph{retention} from \emph{output}. The Pareto elite pool (a Pareto relation without constraint) retains both feasible and nearly-feasible strategies, preserving the ``one-step-away'' genetic information; the recall gate (a hard constraint independent of the retention logic) filters feasible candidates at output, guaranteeing deployment feasibility and confining the verification budget to candidates still in the feasible region. This decoupling is not a simple juxtaposition of Pareto and gate, but a direct response to the property that ``multi-objective + hard-constraint'' cannot be merged into a single solvable problem in text space.

\clearpage
\onecolumn

\section{Global Generation Prompt Template}
\label{sec:prompt-global}

The prompt template used for the initial Skill generation in this framework. The prompt takes the evolution set $\mathcal{D}_{\text{evo}}$ as input and performs three steps in one shot---sample-feature analysis, detection-intent understanding, and Skill design---producing an initial Skill $S_0$ with full-distribution coverage. Placeholders follow Jinja2 syntax; \texttt{\{\{ inference\_template \}\}} denotes the insertion point of the deployment's inference-time prompt template, ensuring that the generated Skill stays compatible with inference-time invocation.

\begin{Verbatim}[fontsize=\footnotesize,breaklines=true,breakanywhere=true,bgcolor=gray!8,frame=single,framerule=0.35pt,framesep=6pt,rulecolor=\color{gray!40}]
# Role
You are a Skill design expert. Based on the analysis of the evolution set,
generate the initial Skill.

# Inputs
1. Inference template {{ inference_template }}
   Deployment-side inference prompt template.
2. Evolution-set path {{ evolution_set_path }}
   CSV in production; readable via shell tools.
3. Available tools {{ available_tools }}
   Read-only tools available during generation.

# Evolution-set analysis steps
Step 1 Sample-feature analysis: count positive/negative ratio; summarize
   common features of each class; distinguish auxiliary noise-reduction
   cues from sufficient risk-exclusion evidence; identify the key
   judgment dimensions separating the two classes.
Step 2 Detection-intent understanding: identify the core objective of
   the detection intent; analyze how it maps to sample features;
   identify boundary scenarios prone to false positives.
Step 3 Skill design: determine core principles; design a structured
   triage strategy; identify empirical lessons and edge cases to
   emphasize.

# Skill generation requirements
1. Clear core principles: 3-5 highest-priority rules.
2. Structured triage strategy: step-by-step analysis framework and
   decision logic.
3. Cover typical scenarios: cover the main patterns of both classes.
4. Unambiguous phrasing: use conditional sentences.
5. Concise and executable: directly guide Agent behavior, avoid
   redundant background.
6. Rigorous triage: distinguish risk-behavior evidence, environmental
   context, risk-exclusion evidence, uncertain information, and
   process/label metadata; "suggest dismiss" must come with a
   risk-exclusion chain.
7. Bounded whitelist-style experience: noise-reduction experience must
   state applicability conditions, non-applicable exclusions, and
   priority when conflicting with risk-behavior evidence.
8. Tool-usage policy: optional, only when information is insufficient
   or verification is needed; results must not be hardcoded into the
   Skill as fixed values or fixed allow/deny lists.
9. No speculation of external knowledge: do not preset trusted
   vendor/domain lists; do not infer external attributes from
   domains/program names alone.

# Recommended Skill structure (Markdown)
- Core principles: 3-5 highest-priority rules.
- Triage strategy: step-by-step reasoning paths and decision logic.
- Empirical lessons: high-value lessons, with applicability boundaries
  and conflict-handling.
- Edge cases: handling of abnormal/under-informed/ambiguous inputs.

# Output format
1. Thought process, wrapped in a ```thought block.
2. Skill content, wrapped in a ```skill block.
\end{Verbatim}

\clearpage

\section{Multi-Round Reflection Prompt Template}
\label{sec:prompt-reflect}

The prompt template used for multi-round reflection in the framework. The prompt takes the current Skill and the Agent's within-batch execution trajectories as input, identifies error patterns, and revises the Skill. It shares the inference-time compatibility constraint, Skill generation requirements, and recommended Skill structure with the global generation prompt (Appendix~\ref{sec:prompt-global}); the differences are mainly in the reflection input and analysis steps.

\begin{Verbatim}[fontsize=\footnotesize,breaklines=true,breakanywhere=true,bgcolor=gray!8,frame=single,framerule=0.35pt,framesep=6pt,rulecolor=\color{gray!40}]
# Role
You are a Skill optimization expert. Based on the current Skill's
execution trajectories and feedback on a batch, revise the Skill to
improve triage accuracy and consistency.

# Inputs
1. Inference template
   {{ inference_template }}
   Deployment-side inference prompt template.
2. Current Skill
   {{ current_skill }}
   The Skill to be revised.
3. Batch trajectories
   {{ batch_trajectories }}
   Each entry contains: sample identifier, alert details, Agent reasoning trajectory, model prediction, feedback.
4. Available tools
   {{ available_tools }}
   Read-only tools available during reflection.

# Analysis principle: no local view
When analyzing error samples, attributing label conflicts, or comparing
similar samples, do not conclude based on "features look similar". Two
similar samples that receive different human dispositions often differ
along other dimensions. Take the operator's perspective, combine the
detection intent, and enumerate dimensions that could explain the
disagreement; find the dimension that truly discriminates them and
encode it as an executable condition in the Skill, rather than simply
lowering the threshold or dismissing outright.

# Analysis steps
Step 1 Error-pattern identification: count errors and proportion;
   summarize common issues (misjudgment conditions, missing key
   information, broken reasoning chains); analyze whether errors
   relate to missing or ambiguous Skill descriptions; check whether
   false-positive fixes harden weak-context features into one-vote
   dismiss conditions.
Step 2 Correct-sample attribution: identify key decision points where
   the model was correct; extract reusable reasoning strategies.
Step 3 Skill gap analysis: compare the current Skill against correct
   sample trajectories; identify what to add, modify, or delete.

# Skill generation requirements
1. Cover error scenarios: provide explicit handling rules for the
   identified error patterns.
2. Reinforce correct strategies: make the high-quality reasoning paths
   from correct samples explicit and followable.
3. Clear core principles: 3-5 highest-priority rules.
4. Structured triage strategy: step-by-step analysis framework and
   decision logic.
5. Cover typical scenarios: cover the main patterns of both classes.
6. Unambiguous phrasing: use conditional sentences.
7. Concise and executable: directly guide Agent behavior, avoid
   redundant background.
8. Rigorous triage: distinguish risk-behavior evidence, environmental
   context, risk-exclusion evidence, uncertain information, and
   process/label metadata; "suggest dismiss" must come with a
   risk-exclusion chain.
9. Bounded whitelist-style experience: state applicability conditions,
   non-applicable exclusions, and priority when conflicting with
   risk-behavior evidence.
10. Tool-usage policy: optional, only when information is insufficient
    or verification is needed; results must not be hardcoded into the
    Skill as fixed values or fixed allow/deny lists.
11. No speculation of external knowledge: do not preset trusted
    vendor/domain lists; do not infer external attributes from
    domains/program names alone.

# Recommended Skill structure (Markdown)
- Core principles: 3-5 highest-priority rules.
- Triage strategy: step-by-step reasoning paths and decision logic.
- Empirical lessons: high-value lessons, with applicability boundaries
  and conflict-handling.
- Edge cases: handling of abnormal/under-informed/ambiguous inputs.

# Output format
1. Thought process, wrapped in a ```thought block.
2. Skill content, wrapped in a ```skill block.
\end{Verbatim}

\clearpage
\twocolumn

\section{Supporting Mechanism Details}
\label{sec:supporting}

This appendix gives the formal definitions of the supporting mechanisms (scoring weights and the Pareto elite pool). The design rationale for the other mechanisms (roulette parent selection, the per-parent cap $\kappa$) is already covered in the main text.

\subsection{Scoring Weights}

The cost structure of alert triage is inherently asymmetric (as stated in the problem formulation): missed threats (FN) cost far more than false positives retained in the human queue (FP). Elkan~\cite{elkan2001foundations} established that the optimal decision threshold in such asymmetric-cost settings is determined by the cost ratio, and Lee and Stolfo~\cite{lee2002toward} further introduced cost-sensitive modeling into intrusion detection. REFINE assigns asymmetric weights to each prediction outcome---TP (correct escalation) $= +1$, TN (correct dismissal) $= +3$, FN (missed threat) $= -1$, FP (false alarm kept in queue) $= 0$---forming the scalar $\mathrm{score}(S, \mathcal{D})$. The gate guarantees $\text{FN}=0$ at output, so scoring among feasible strategies reduces to a linear function $w_{\text{TN}} \cdot |\text{TN}| + w_{\text{TP}} \cdot |\text{TP}| + w_{\text{FP}} \cdot |\text{FP}|$ that directly maximizes the alert takeover rate; any weight combination satisfying $w_{\text{TN}} > w_{\text{TP}} > w_{\text{FP}} \geq 0$ produces the same strategy ranking. Scoring determines parent-selection probability and the success-constraint judgment; it does not directly decide which Skill enters the elite pool---that is governed by the Pareto relation below.

\subsection{Formal Pareto Elite Pool Definition}

REFINE adopts GEPA's~\cite{agrawal2026gepa} \emph{per-instance} non-dominated retention: each evolution-set sample $x$ is treated as a separate objective. With the per-sample score $s(S, x)$ (from the weights above), the champions of $x$ are $\mathcal{C}(x) = \arg\max_{S \in \mathcal{S}} s(S, x)$, i.e., the candidates handling $x$ correctly; the elite pool is their union $\mathcal{P} = \bigcup_{x \in \mathcal{D}_{\text{evo}}} \mathcal{C}(x)$, after pruning any candidate whose champion set $\{x : S \in \mathcal{C}(x)\}$ is a strict subset of another's. Retaining any strategy that is uniquely best on even one sample---though globally lower in accuracy or recall---preserves the intermediate diversity that a coarser $(\text{accuracy}, \text{recall})$ dominance would discard.


\end{document}